\documentclass[preprint,10pt]{aastex} 
\usepackage{amsmath}
\usepackage{amssymb}
\usepackage{wasysym}

\usepackage{color}
\newcommand{\code}[1]{\texttt{#1}}
\newcommand{\eps}{\varepsilon}
\newcommand{\about}{\sim\!}
\newcommand{\sub}[1]{_{\text{#1}}} 
\newcommand{\unit}[1]{\ \mathrm{#1}} 

\usepackage[breaklinks=true,colorlinks=true,citecolor=blue]{hyperref}
\usepackage[all]{hypcap}

\begin{document}

\title{Numerical modeling of the disruption of Comet D/1993 F2 Shoemaker-Levy 9 representing the progenitor by a gravitationally bound assemblage of randomly shaped polyhedra}
\author{Naor Movshovitz}
\email{nmovshov@ucsc.edu}
\author{Erik Asphaug}
\and
\author{Donald Korycansky}
\affil{Department of Earth and Planetary Sciences, University of California, Santa Cruz,\\ 1156 High Street, Santa Cruz CA, 95064}

\begin{abstract}
We advance the modeling of rubble-pile solid bodies by re-examining the tidal breakup of comet Shoemaker-Levy 9, an event that occurred during a $1.33\ R_{\jupiter}$ encounter with Jupiter in July 1992. 
Tidal disruption of the comet nucleus led to a chain of  sub-nuclei $\sim 100-1000$ m diameter; these went on to collide with the planet two years later 
\citep{Chodas1996}. They were  intensively studied  prior to and during the collisions, making SL9 the best natural benchmark 
for physical models of small body disruption.  For the first time in the study of this event, we use  numerical codes treating rubble piles as collections of 
polyhedra \citep{Korycansky2009}. This introduces forces of dilatation and friction, and inelastic response.  As in our previous studies \citep{Asphaug1994,Asphaug1996} we conclude that the progenitor must have been a 
rubble pile, and we obtain approximately the same  pre-breakup diameter ($\about{1.5\unit{km}}$) in our best fits to the data.  We find that the inclusion of realistic fragment 
shapes leads to grain locking and dilatancy, so that even in the absence of friction or other dissipation we find that disruption is overall more difficult than in our spheres-based simulations. We constrain the comet's bulk density at $\rho_{bulk} \about{300-400}\unit{kg\;m^{-3}}$, half that of our spheres-based predictions and consistent with recent estimates derived from spacecraft observations. 
\end{abstract}



\section{Introduction}\label{sec:intro}
\begin{figure}[t]
\label{fig:hst}
\centering
\includegraphics[width=\textwidth]{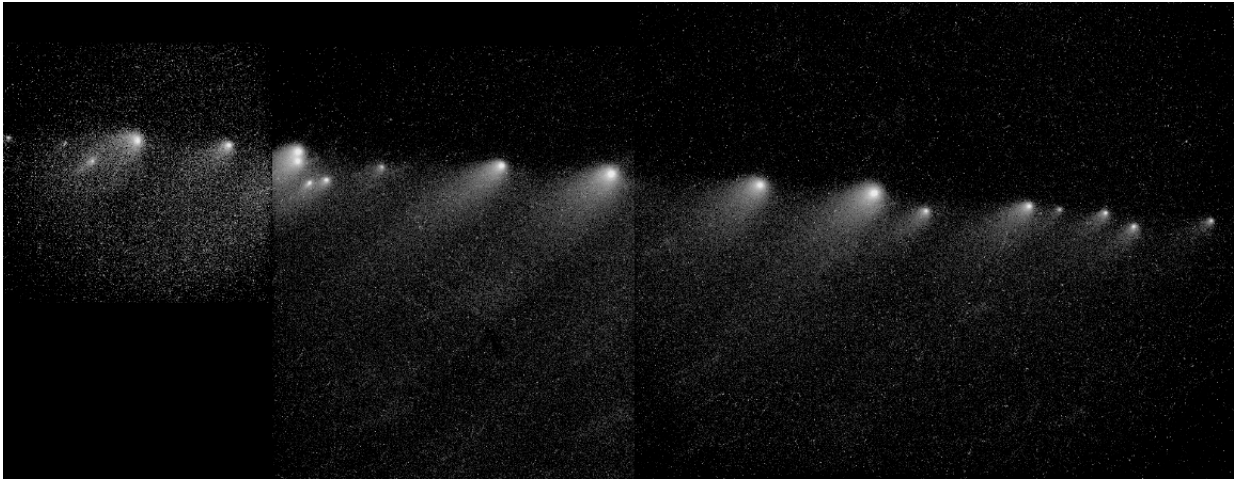}
\caption{Hubble Space Telescope image \citep{Weaver1995} showing the `string of pearls' comet Shoemaker-Levy 9 in March, 1994, four months before its collision with the planet Jupiter.}
\label{fig:sl9real}
\end{figure}

Most kilometer-sized asteroids are
likely rubble piles \citep{Richardson2002,Fujiwara2006,Asphaug2009a}. Many comets may also be strengthless
or nearly strengthless bodies, their fragility demonstrated
when they break up far from perihelion for no
obvious reason \citep{Weissman2004}. One comet was observed shortly after
it broke up for a very obvious reason: Comet Shoemaker-Levy 9 (SL9) made a spectacular plunge into
Jupiter following a close approach two years previously,
that tidally disrupted the original progenitor into at least
21 detectable pieces (Fig.~\ref{fig:hst}), as summarized in \citet{Noll1996}. 

Tidal disruption by Jupiter is not unusual among the population of short-period comets. Numerous crater chains on Ganymede and
Callisto \citep{Schenk1996} provide direct evidence for $\sim 10$ events where comets disrupted by Jupiter slammed into one of the Galilean satellites. For each such imprint, many thousands of disrupted comets went on to become families of bodies.  Similarly, comets coming too close to the Sun are tidally disrupted, the most famous
being the Kreutz family of Sun-grazers studied by \cite{Marsden1967} and \cite{Knight2010}. Other planets also disrupt comets, although Saturn's density may be too
low to lead to tidal disruption of bodies of cometary density without resulting in a collision.  Asteroids, of higher density than comets, can be 
tidally disrupted as well, but only in encounters with correspondingly denser terrestrial planets.
If we regard SL9 as a typical case, we may use the details of its final orbit to infer something about the
 physical properties of comets in general, and more broadly, about the physics of rubble piles.  
 We are also studying the Kreutz family in a similar level of detail, (Weissman et al., in preparation).
 
 \subsection{Models of Tidal Disruption}

A number of studies, conducted before and after the discovery of SL9, looked at the details of tidal distortion and disruption. Most directly applicable was the detailed analytical study  by \citet{Sridhar1992} who analyzed the tidal stresses exerted on a homogeneous, incompressible, viscous fluid as it passes by a planet on a parabolic trajectory. Assuming that the body remained ellipsoidal during deformation, they concluded that an orbit that results in mass shedding (that is, gravitational instability of the ellipsoid) must have pericenter distance $Q<1.05(M\sub{p}/\rho_0)^{1/3}$, where $M\sub{p}$ is the mass of the planet and $\rho_0$ is the fluid (planetesimal) density.

\citet{Asphaug1996} tested, extensively, the possible outcomes of a tidal encounter with Jupiter, for a small body of varying strength and density. They started by showing, using an SPH-based hydrocode with strength \citep{Benz1994,Benz1995}, that a solid body, no matter how weak, can be ruled out as the progenitor SL9. A solid body was found to break in half, as soon as the tidal stress was able to activate the weakest flaw and propagate a fracture across the body. The stress is thus relieved, and must build up again to an even higher value in the two resulting fragments, which have a higher failure threshold, and have a tidal stress reduced owing to their smaller size. On a nearly parabolic orbit, such as was deduced for SL9, this process is too slow to repeat more than once or twice, no matter how low the initial strength, and so is inconsistent with the $>20$ fragments observed for SL9. 

\citet{Asphaug1996} then proceeded to test, using a particle code with elastic collisions, the possible outcomes when the progenitor is instead a rubble pile, with varying density. They first tested their rubble pile (or `marble pile') code against the analytical predictions of \citet{Sridhar1992} and found good agreement for the threshold of mass loss for incompressible fluid bodies on parabolic encounters.  With some confidence in the physics of their code based on this good analytical match, they were then able to constrain a bulk density for the progenitor ($\rho\approx{600}\unit{kg\;m^{-3}}$) from the resulting chain morphology (number and distribution of fragments). This was well within the allowable range predicted by \citet{Boss1994}, who predicted $\rho\lesssim {1100-2400}\unit{kg\;m^{-3}}$.  They were then able to constrain a diameter ($d\approx{1.5}\unit{km}$) from the chain length. They were able to test the effect of friction, in a very simple approach where they froze all relative grain motions until the time of periapse, at which time if the comet was going to disrupt, this would be the moment of peak stress, after which the modeled comet nucleus was again treated as a  pile of frictionless spheres.  The corresponding sizes and masses of fragments resulted in good agreement with the best fits to impact plume thermal characteristics observed by the Galileo spacecraft (which was at the time en route to Jupiter, as luck would have it) and modeled using the CTH hydrocode \citep{Crawford1995}. 

Earlier studies \citep{Asphaug1994,Solem1994} applied similar models of cometary rubble piles, assuming elastic spheres bound
 together by only gravity, to derive a density and a diameter for SL9.  Both groups obtained similar diameters and densities for the comet.   \cite{Asphaug1996} and  \cite{Schenk1996}~furthermore showed that a rubble 
-pile structure is the only possible explanation to SL9 that also fits the data of cometary disruption remnants imprinted
 upon the surfaces of Ganymede and Callisto.  (Io and Europa have surfaces that are too young to record `tidal catenae', the chains of craters formed by a tidally disrupted progenitor.)  

Similar studies to constrain size, density and internal structure have been conducted for Near-Earth Asteroids (NEAs). \citet{Solem1996} showed how a rubble pile of spherical ``rocks'', similar to the marble-piles just described, can deform to make an elongated post-encounter shape. \cite{Bottke1996a} approximated a rubble pile asteroid as a rotating contact-binary (that is, two contacting spheres) and found that it takes one or more tidal encounters before impact to sufficiently separate the fragments, if these are to be the source of doublet craters on Earth or on Venus.  This paper made the successful prediction that about one-sixth of all near-Earth asteroids are well separated binaries \citep{Weidenschilling1989,Merline2002} in proportion to the fraction of primary impact craters that are doublets.

\cite{Richardson1998} used the \code{pkdgrav} N-body particle code to make a thorough investigation of the possible outcomes of a close tidal encounter of a rubble-pile asteroid with the Earth. Their code, discussed further below, represented collisions somewhat more realistically, by including a tangential coefficient of restitution approximating friction, in addition to the normal coefficient of restitution.  By considering elongated, rotating progenitors on a hyperbolic trajectory, they were able to peek into three previously closed dimensions of the parameter space. They showed how an elongated progenitor with prograde rotation is far easier to disrupt than a spherical, non-rotating one. Indeed one of their classes of outcomes was designated S-class, for ``SL9-type'' disruption.

In recent studies, \citet{Holsapple2006,Holsapple2008} have extended the work of \cite{Roche1847} and \cite{Jeans1917}, removing the assumptions of a fluid body with a specific axes ratio and considering instead solid, spinning, ellipsoidal bodies subjected to tidal forces. Using a Drucker-Prager strength model with zero cohesion, parametrized by an angle of friction, they find limit distances for total failure for a long list of special cases, including the case of a stray asteroid passing by a primary. They find that bodies with even a moderate amount of strength can venture much closer to a planet than the classical Roche limit predicts. For example, with an angle of friction $\phi=30^\circ$ a passing prolate body with aspect ratio $\alpha=\beta=0.6$ and no spin could approach as close as $d=1.53R(\rho\sub{P}/\rho)^{1/3}$ without disruption, $R$ being the radius of the primary and $\rho\sub{P},\rho$ the densities of the primary and passing body, respectively. Being a static theory, however, the \citeauthor{Holsapple2006} does not include a possible spin-up of the asteroid as it passes the primary, or the dynamics of disruption when it does occur.

\subsection {Asteroids and Comets}

Two possible tidal catenae are identified on the Moon \citep{Melosh1994}, which would have been caused by the tidal breakup of NEOs (Near Earth Asteroids) around the Earth, crashing into the Moon after a voyage through space of some 30--60 Earth radii depending on whether the collisions happened early or late \citep{Bottke1997}.  They occur, as they must, on the Earth-facing hemisphere.  In recent LROC images of the Moon there are additional possible examples of smaller catenae that appear to have tidal disruption morphologies, that would be caused by progenitors hundreds of meters across -- the size of the rubble-pile, Earth-crossing asteroid 25143 Itokawa \citep{Fujiwara2006}. It is, however, quite challenging to discern tidal catenae from garden-variety catenae that result from the far-flung secondary cratering ejecta common on the Moon; none of these smaller features have yet been validated or reported in the literature, and we remain agnostic.  

The possibility of NEO-derived catenae on the Moon is a vital consideration, for if the Moon bears a record of kilometer-sized and smaller tidally disrupted NEOs, just as Ganymede and Callisto bear a record of tidally disrupted Jupiter-family comets \citep{Schenk1996}, then we have the opportunity to learn NEO structural and mechanical characteristics in a manner that can allow for more informed planning of mitigation strategies for potentially hazardous NEOs \citep{NRC2010}.  Until then SL9 is arguably a reasonable proxy for studying rubble-pile asteroids, and perhaps  the only available proxy. Asteroids are denser than comets, and Earth is denser than Jupiter by about the same fraction.  So the key parameter being the Roche limit, these situations are not dissimilar, to the extent that either body can be represented as aggregates of polyhedra with specified coefficients of friction and restitution (as described below).  

So long as the periapse distance is normalized to the Roche limit, 
\begin{equation}
R\sub{roche}=2.44\;R\sub{planet}(\rho\sub{interloper}/\rho\sub{planet})^{1/3},
\end{equation}
and the encounter velocity normalized to the escape velocity from the planet at the Roche distance, model outcomes such as those reported by  \cite{Asphaug1996} are scale invariant.  That is, asteroids passing near the Earth are similar to comets passing near Jupiter (e.g.~$\rho\sub{interloper}/\rho\sub{planet}\approx{0.3}$ for each), and a 1~km rubble pile passing near Jupiter looks the same as a 100 km rubble pile on the same orbit, if the yardstick is 100 times as big.  Scale invariance begins to break down when 
stress dependent parameters such as friction are introduced.  

In that sense, the study of SL9 is a rather general study.  Of key importance to making further scientific progress in the area of understanding the geophysics of small bodies, the \emph{techniques} used for studying rubble-pile comets and rubble-pile asteroids are identical. What we learn from NEA simulations is useful when we turn to look at Jupiter crossing comets, and we  expect that what we learn in this study of a Jupiter family comet to be applicable to any catenae that may be found on the Moon.

Concerning mitigation of hazardous comets and asteroids \citep{Belton2004} a typical technique is to use an impact or explosion 
to divert or destroy a body perhaps $\sim 100-1000$ m diameter. This involves accelerating material to at least a substantial fraction of the escape velocity of the target body.  On Earth, this is 11.2 kilometers per second, much faster than any of the ejecta, so  events unfold to completion quickly at the scales of cratering that can be observed in the laboratory or in the field.  But on an SL9-sized body, escape velocity is about one meter per second, and the dynamical (self-gravitational) timescale is hours. The opening time of a disruptive collision is also measured in hours, and the speed is comparable to a low-velocity landslide -- analogous in many respects to tidal disruption.  Thus what we learn about tidal disruption, is expected to be applicable generally to crater formation and disruption of comets and asteroids and NEOs.

\subsection{From Marble Piles to Rubble Piles}

One common feature of previous dynamical studies of SL9, and of Earth tidal encounters by rubble-pile NEOs, is the use of hard spheres as the building blocks of the simulated rubble pile. Considering the computational capabilities of the time of the Shoemaker-Levy 9 impacts, this was a necessity of being able to model the event at all in a 3D numerical framework: treating the components as radially symmetric and with a simple restitutive potential that balanced self-gravity at resting contact. But this approach does not fully capture the ability of a rubble pile to withstand shear stresses.

When a granular material is under shear stress, the maximum allowable stress before some failure occurs is a consequence of the interlocking of granular particles. This ``strength'' of the granular material is often assumed to be proportional to the confining pressure, since it results from the interlocking particles having to move each other out of the way, working against the confining pressure. The linear relationship between maximum shear stress and confining pressure is often characterized using a \emph{friction angle} parameter, and it is known, for example, that closely packed, uniform, rigid, frictionless spheres support a friction angle (note the somewhat inappropriate name) of about $23^\circ$ \citep{Albert1997}. It is also known that this friction angle can be greatly modified (usually lowered) by using a size distribution of spheres, and it could be the case that materials composed of rough, non-spherical grains will also exhibit a different angle of friction, and thus a different response to tidal shearing.

A way to parametrize the effect of interlocking grains while still using spherical elements is to use a \emph{soft sphere} method, such as implemented recently by \cite{Schwartz2012}. In this method a spring and dashpot force is applied to interacting grains if they experience lateral relative motion, and the strength of the spring can be adjusted to mimic the desired friction angle. We chose to take a complimentary approach, and simulate the interlocking directly by modeling a rubble pile with polyhedral grains. Polyhedral grains are expected to display different (likely higher) friction angles, since even rotating a single grain, in place, requires work to be done against the confining pressure. Simulating polyhedral grains, however, is considerably more difficult and computationally expensive than simulating contacts between spheres only. On the up side, our approach avoids the use of very small time steps that are the hallmark of soft sphere methods.

In the present work we revisit the question of SL9's disruption with much improved computational capabilities and simulation methods. Our logic and procedure are very similar to the studies by \cite{Asphaug1996} and \cite{Richardson1998}, but with the advantage of faster machines and more versatile algorithms, including public domain and commercial codes originally developed for computer games, that are very well adaptable to the  physics of rubble-pile collisions \citep{Korycansky2009}.  Our goal, as was theirs, is to constrain SL9's structure, diameter and density, and thereby to gain a fundamental understanding of the physics of comets. Below we describe in detail our procedure and the numerical tools we employ. We then present the results of several sets of runs, and argue that these results suggest a value of $\rho\apprle{300-400\unit{kg\;m^{-3}}}$ for SL9's bulk density.

But first, we must pause to entertain the notion that this is not the final answer to be obtained by N-body methods -- that in another fifteen years, another paper will come along with still better methods, and claim a density half again smaller for SL9,  defending our present attempts as what one would do with the crude technology of the early 2010s.  We do not believe this to be the case, but we wish to emphasize the importance of finding benchmarks for these kinds of codes at every opportunity, including in space station experiments and other stable microgravity research platforms.  The limitation to spherical particles was a fundamental one which is now lifted.  Given that our results are only weakly dependent upon friction and restitution (see below), it appears that the major distinction between the models of the 1990s and our present models is the aspherical geometry, leading to grain stacking and dilatancy that limits grain motion.  The next steps in granular modeling will be in even more detailed irregular shapes (e.g.~with fractal-like rather than planar surfaces) and the inclusion of cohesion forces.  However we can argue (though not perhaps convincingly, even to ourselves) that fractal-like surfaces, i.e.~roughness, are already well represented in the planar friction model, and that cohesion was already ruled out on the basis of the `string of pearls' morphology of the post-disruption SL9, indicative of a self-gravity dominated process \citep{Chandrasekhar1961,Hahn1998a}, and in the scale-invariance of the catenae on Ganymede and Callisto \citep{Schenk1996}.  But granular physics is still a young science and there may yet be devils in the details.  

\section{Method}
\subsection{Discrete Element Models}
We employ DEM models using randomly shaped polyhedra as building blocks of the simulated rubble pile, with resolution much higher than previously applied to this problem. By using these polyhedral ``grains'' our models are attempting to directly simulate the cause of granular friction. As mentioned, \cite{Asphaug1996} represented a maximal friction by not allowing any inter granular movement until the comet nucleus had arrived at periapse, on the premise that sliding friction is much lower than static friction.  However the shapes of grains require work to be done in moving adjacent grains out of the way, in order to rotate, so that continuum shear is resisted by a force that may be called \emph{dilatation}, to distinguish it from the Coulomb friction applied at the surface of the grains\footnote{It may be argued that the microscopic origin of Coulomb friction is simply the same dilatation effect at work on a smaller scale. In this work we use the term \emph{dilatation} when the effect is the direct result of our very macroscopic grains interlocking. The use of polyhedral grains with flat faces also opens the possibility that grains will slide across each other much as a rectangular box slides down a sloped floor. In our models this is treated by applying a friction force to the surface of the interacting particles, parametrized by a coefficient of friction. Since these two effects are not easily distinguishable in our model, and since both are linearly proportional to the confining pressure, we often refer to the combined effect resisting shear as friction.}. This force can only be directly simulated with non-spherical grains, as spherical grains are free to roll in place without resistance, unless this resistance is parametrized and explicitly supplied by the code, as in soft sphere methods.

It is easy to understand why we obtain, when all is said and done, a lower density than before. Friction resists the tidal shear stress, in a sense substituting for density (self-gravity) as the force that is holding the comet together against disruption. Friction being a short range force, and gravity being a long range force, their interplay defines the final morphology of the fragment chain, and thus we can still obtain a unique solution, even if the effect of friction influences our choice of bulk density.  

The main difficulty when moving from a spheres-based simulation to more complex shapes is the implementation of a fast contact detection algorithm. The direct approach of solving the geometric equations for each side, vertex, and face of every element is out of the question. Fortunately there are better algorithms. GJK \citep{Gilbert1988} is perhaps the most popular as it operates in linear complexity (with the number of vertices for two convex objects) and can handle collisions between curved and polyhedral shapes. The challenge then becomes implementing the geometric algorithm in the most efficient way. As rough shapes necessarily have many vertices, even a linear algorithm quickly becomes too slow when too many shapes are compared against each other. Collision detection must be followed by collision resolution (that is, physical treatment of the collision as before-and-after states). Although this is a well studied part of classical mechanics \citep[e.g.,][]{MacMillan36} it is again not simple to implement an optimal algorithm that can handle collisions (with and without friction) and resting contacts with minimum operations.

Fortunately, a vast amount of research into optimal collision detection and resolution has come out of the computer games industry. These algorithms are the main components of what the video game industry calls \emph{physics engines}. Physics engines are at their core DEM simulations, and the financial incentive of the industry has resulted in the development of a number of reasonably accurate, but \emph{very} fast implementations. The risk in adopting such physics engines is that what is good enough for a game, may not be good enough for science.  On the other hand, a number of these engines have been found to be excellent in their accurate physical treatment of elastic collisions between objects with complex shapes \citep{Longshaw2009}. In the present work we have used for the most part NVIDIA's  \code{PhysX} engine\footnote{http://developer.nvidia.com/physx}, comparing it in key instances with the previous model of \cite{Korycansky2009} utilizing the publicly available Open Dynamics Engine.

The \code{PhysX} engine combines rigid body dynamics and collision detection (and many other potentially useful features) in a single library, making it very easy to use as a basis for a DEM simulation. Among other rigid-body libraries, it stands out in terms of performance and stability. This high performance comes at a cost, however. \code{PhysX} uses single-precision floating-point numbers, and a first-order forward Euler scheme for dynamic integration. Being a closed source product (although free to use and distribute in binary form) these and other choices are not easily modified without acquisition of the source code or direct collaboration with its developers. What single precision, first order solution means in practice, is that we must be careful to chose small enough time steps for integration, much smaller than used in sophisticated state-of-the-art N-body integrators.  We determine the appropriate time step by careful analysis of conservation in test simulations.

\subsection{Validation}

Over the past two years we have validated the correct behavior of the \code{PhysX} rigid-body module in simple and complex, dynamic and static scenarios. To validate the elementary rigid body physics, we performed a suite of fairly simple tests. A rotating-tumbling body with a full moment of inertia tensor was used to verify the correct integration of Euler's equations.  A point mass orbiting in a central gravity potential verified the force and acceleration integrator, but more than 1000 time steps per orbit were required to maintain better than one percent conservation of orbital energy. Note that this is much worse than is possible with higher order integrators.  

The planar friction model was validated by simulating a box sliding on an inclined plane.  The box begins sliding at an incline angle $\theta$ where $\tan{\theta}=\mu$, the friction coefficient, and this behavior is independent of the acceleration of gravity as long as a time step is chosen that is not short enough to underflow the velocity change. A bouncing ball verified the conservation of energy and the correct handling of a restitution coefficient, $\eps$, the ratio of relative speed of separation to the relative speed of impact,  with the exception of \emph{perfectly} elastic bodies ($\eps=1$) that can lead to unstable behavior over time.  

Conservation of linear and angular momentum during collisions was also validated. Here two bodies with random shapes and sizes are set on a collision course and linear and angular momentum are measured before and after collision. Linear momentum is always conserved accurately.
As for angular momentum, we found that  bodies with simple, smooth shapes conserve angular momentum very accurately.  However, the binary collision between two irregular, polyhedral shapes can lead to a change of the binary system's angular momentum vector by up to 30\% in magnitude, perhaps corresponding to an approximation made in the collision detection.  This is disappointing, and is possibly the result of an approximation made in the interest of speeding the engine. The error, however, is randomly oriented, so that an ensemble of many bodies always conserves its total angular momentum very well. 

This led to tests of more complex physical systems involving aggregates of bodies.  We used \code{PhysX} to simulate an avalanche of polyherdal dice, beginning with a simple laboratory simulation where we let a tall pile of $1400$ 12-sided dice (the kind familiar to a different kind of gamer) collapse, in 1g, to a stable slope in a transparent rectangular container.  We measured the average slope angle with a ruler. This process was then simulated with \code{PhysX}, for the same boundary geometry, and we found an excellent agreement between the measured and modeled angles of stable slope. This is particularly important because the slope angle depends on the correct application of sliding friction and resting contacts between grains.

We have also performed several of our simulations using a second implementation of a DEM, based on the Open Dynamics Engine \citep{Korycansky2009}, a DEM that uses different algorithms, and the results from both codes are always similar, although \code{PhysX} is an order of magnitude faster, thus allowing for much higher resolution simulations. 

As in \citep{Asphaug1996} we do not hope to resolve every actual granule in a cometary nucleus.  At best, our grains are tens of meters across, representing an assemblage of much smaller particles.  Even with billions of grains, we would simulate SL9 with at best meter-sized objects, when in fact comets may be composed of dust and ice.  We merely hope to resolve the granular behavior at a scale smaller than the deformation occurring within the tidally disrupting cometary nucleus.  \cite{Asphaug1996} found similar behavior, with increasing $N$, once the comet was resolved with more than several hundreds of spheres.  We likewise find similar behavior, but converging at higher `resolution' (several 1000s polyhedral grains), owing to the fact that the grain-grain interactions are more complex. We note, however, that increasing the number of grains in a simulation will be necessary to test the effects of different grain size distributions, another potentially important dimension of the parameter space, not explored here.

We caution that `resolution' is not really the appropriate word here, because even in the limit of infinite numbers of grains we are not resolving a continuum.  Indeed at infinitesimal grain diameter the material would have infinite friction. Rather, each simulation, for a given $N$, is actually a somewhat different physical system. But as long as we have enough grains to simulate the inter granular effects such as dilatancy and stacking, and the frictional forces, further increases in $N$ do not make any apparent difference in our primary scientific result, the determination of the density of SL9, or in the physical and morphological behavior of the comet's tidal disruption.  Insofar as we can adequately capture shear friction in a granular mass, and insofar as we can capture the physics of dilatancy caused when rubble expands during shear (requiring energy), the approach is likely to capture the most significant physical aspects of the tidal disruption of rubble piles.

\subsection{Simulating tidal encounters}
A rubble pile is simulated in a DEM by a collection of rigid body elements we call \emph{grains}. We make a rubble pile by letting a collection of randomly shaped polyhedra collapse freely into a roughly spherical aggregate. Making a kilometer sized body out of several thousands elements, means that each element is $\about{100\unit{m}}$ in size. (We use a uniform, narrow size distribution, intended only to inject more randomness to the grain assemblage. The smallest grain has a bounding sphere about half the diameter of the largest grain's bounding sphere.) An actual cometary rubble pile is surely made of a wide distribution of grain sizes, with the near-surface grains likely much smaller than a meter in the case of comets \citep[based on imaging and thermal inertia data, e.g.][]{Fujiwara2006}. Although this limitation is important to keep in mind,  the essential physics of the process we investigate, namely granular shear flow, presents itself at these larger grain sizes.  Models at successively higher resolution plateau to a common result.  

When the ``comet'' is ready, we send it on a pre-calculated orbit past Jupiter. The orbit of SL9 was an almost parabolic orbit with eccentricity $e=0.997$ and perijove $q=1.33$ Jupiter radii \citep{Sekanina1994}. To allow for pre-encounter tides, and to follow it out to dynamical completion, we calculate the comet's orbit between four Roche limits pre-perijove, and $15$ Jupiter radii post-perijove. Thus the simulation starts before any significant tidal force is felt by the comet, and continues past the orbit of Ganymede. The orbit is divided into equal time steps of one thousandth of the dynamical time, $dt=10^{-3}(2\rho{G})^{-1/2}$, where $G$ is the gravitational constant and $\rho$ stands for the comet's average density, not the material density of grains. The average density is in fact the main control parameter of the simulation, and we set it by adjusting the individual grain masses after determining the approximate volume of the body.

For reasons of numerical accuracy, the simulation is done in the frame of reference of the comet. In every time step, the net force on each grain is calculated by adding the gravitational attraction from all other grains (computed pairwise), the gravitational attraction of Jupiter, whose position is read from the pre-calculated orbit, and the fictitious force from the non-inertial comet-centered frame. These external forces are applied at each time step to every grain, while the accelerations resulting from collision between grains, as well as the time integration, are handled by the \code{PhysX} library.  As mentioned, the conservation of energy and momentum are validated as described above, as well as the friction forces, and angular momentum is validated for interactions averaged over assemblages of grains.

\begin{figure}
\centering
\includegraphics[width=\textwidth]{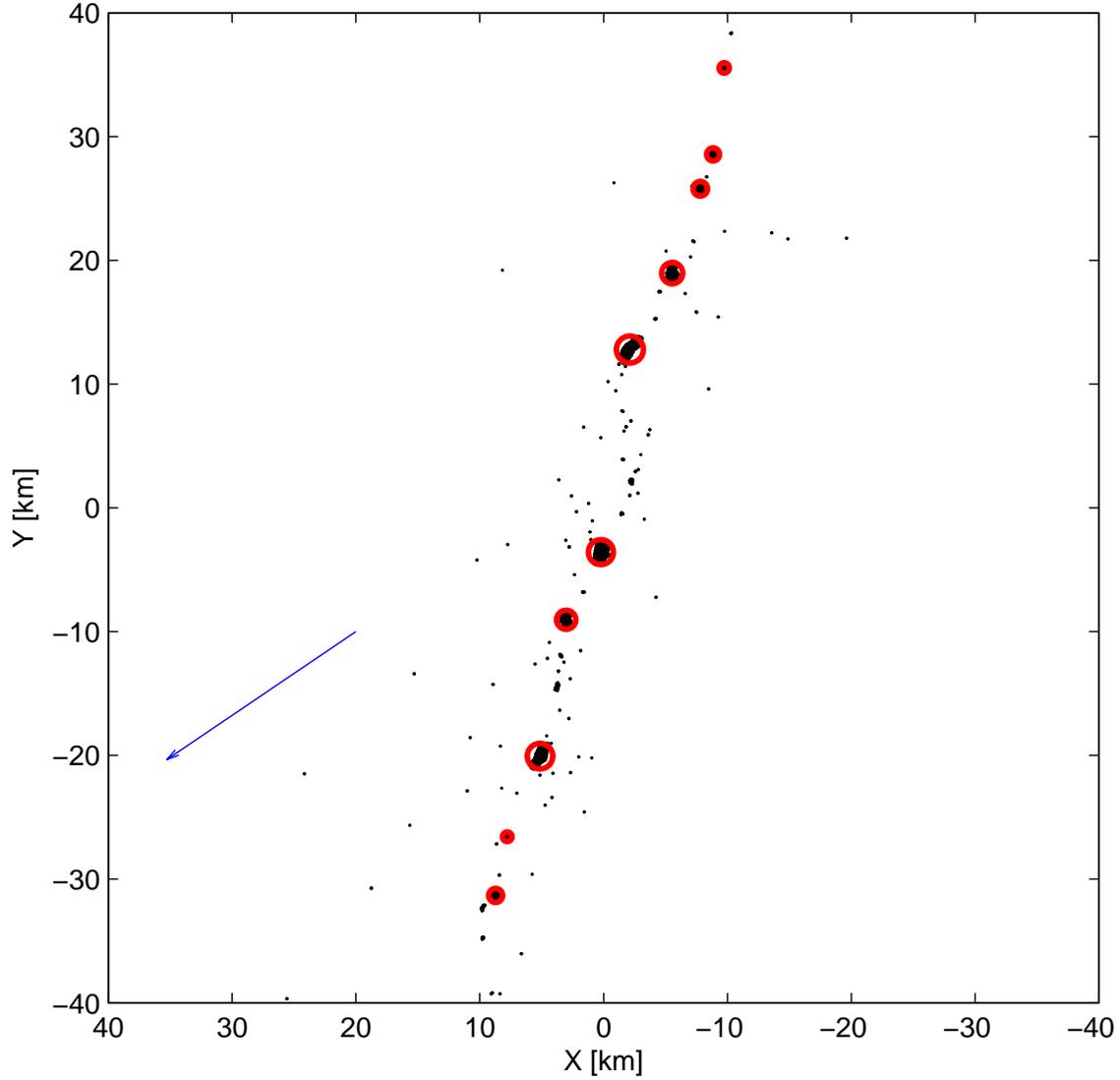}
\caption{Tidal break-up of a simulated comet, shown a few hours after perijove. The projection on the plane of the orbit of each grain is marked by black dots. Thick circles mark some of the larger fragments, the largest of which contains about $19\%$ of the original mass. The arrow points to the position of Jupiter, about $10^6\unit{km}$ away. This run was made with a bulk density $\rho=300\unit{kg/m^3}$.}
\label{fig:sl9}
\end{figure}

The result of a typical test run with 4096 grains is shown in figure~\ref{fig:sl9}. The position of each grain is plotted in the plane of the orbit (the out-of-plane extent of the fragment train is relatively small). Red circles mark clusters of grains, somewhat subjectively defined as groups of touching or nearly touching grains with at least one per-cent of the mass of the initial body. (The red circles are just a visual aid, they have no physical meaning.) By this stage the clustering had completed, and the fragments continue to separate at a rate determined by simple, two-body orbital dynamics. (Non-interacting particles along the same parabolic orbit separate at a rate proportional to a $4/3$ power of time \setcitestyle{square}\citep{Sridhar1992}\setcitestyle{round}.) Once beyond the Roche limit of the planet, for appropriate bulk density rubble piles we see the signature SL9 `string of pearls,' i.e., a well separated, roughly linear train of more than a few, roughly equal sized fragments.  The self-gravitational clumping instability that forms these new cometary nuclei was described by \citet{Chandrasekhar1961} and by \citet{Hahn1998a}. 

This clumping is a sensitive measure of progenitor density.  Starting with an over dense progenitor, the outcome would instead of the above figure, be one or two large fragments containing almost all the mass of the comet. Conversely, if the progenitor's density was too low, no clustering would occur and the comet would ``atomize'' into individual boulders, gravels and dust.  Because we know the encounter orbit so precisely \citep{Chodas1996} it is possible to constrain the average density of SL9 in this manner. Although the classification of a given outcome is somewhat subjective (too few fragments, too many fragments, or just right) the two end members are easily and unambiguously spotted (see Sec.~\ref{sec:mainresult}), leaving a narrow range of possible densities inbetween.

\section{Results}
\subsection{Simulation parameters}
The main quantity we are interested in constraining with these simulations is the bulk density of the progenitor SL9 comet. Several other physical parameters, however, are likely to influence a tidal encounter. The most important are discussed next.

\subsubsection{Friction coefficient} 
\begin{figure}
\centering
\includegraphics[width=\textwidth]{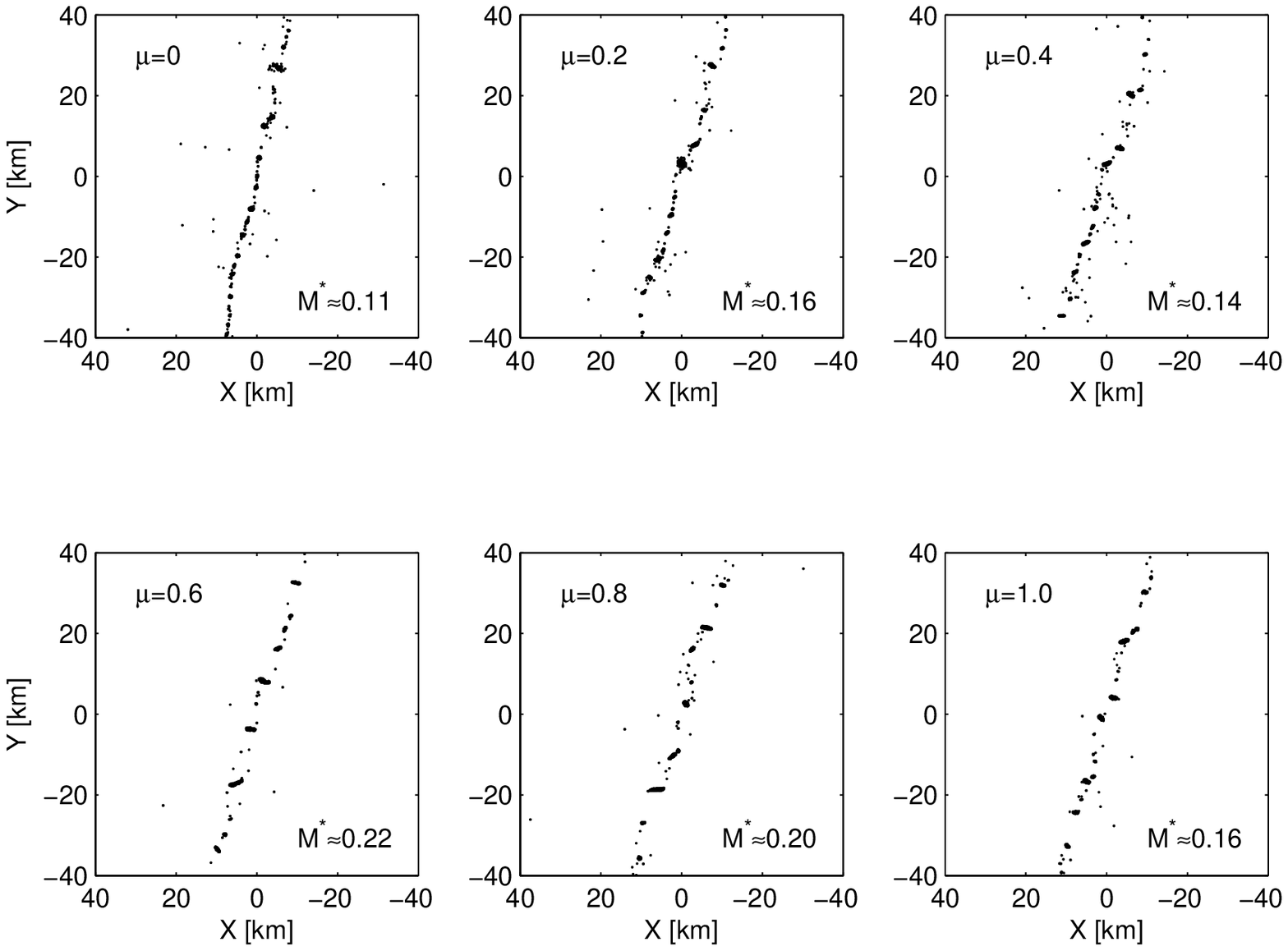}
\caption{Projection to the plane of orbit of grains from tidally disrupted rubble piles. All runs followed the same orbit, and started as the same progenitor rubble pile, with a bulk density of $\rho=300\unit{kg/m^3}$. The mass fraction contained in the largest clump is indicated by $M^*$ (but note that our definition of ``clump'' is somewhat subjective). Friction is clearly a significant force, helping to delay breakup until tidal shear exceeds some critical value.}
\label{fig:fric2}
\end{figure}

Although including \emph{some} inter-granular friction in the simulation is crucial, and easily noticeable, our tests indicate that the exact value of the friction coefficient does not affect the outcome of a given tidal encounter very strongly (Fig.~\ref{fig:fric2}). This is fortunate, since the exact value of the friction coefficient between cometary material grains in vacuum is not well known. However, many measurements are obtained, especially in engineering applications of sea ice, concrete and other materials, where the effective coefficient of friction is of order $0.5-0.8$.  We therefore use a friction coefficient of $\mu=0.5$, typical of ices, rocks and other geological materials, in most of our runs, recognizing that it is possibly a lower limit, with only minor changes for increasing friction in the range of known values for common geologic materials.

\subsubsection{Restitution coefficient} 
\begin{figure}
\centering
\includegraphics[width=\textwidth]{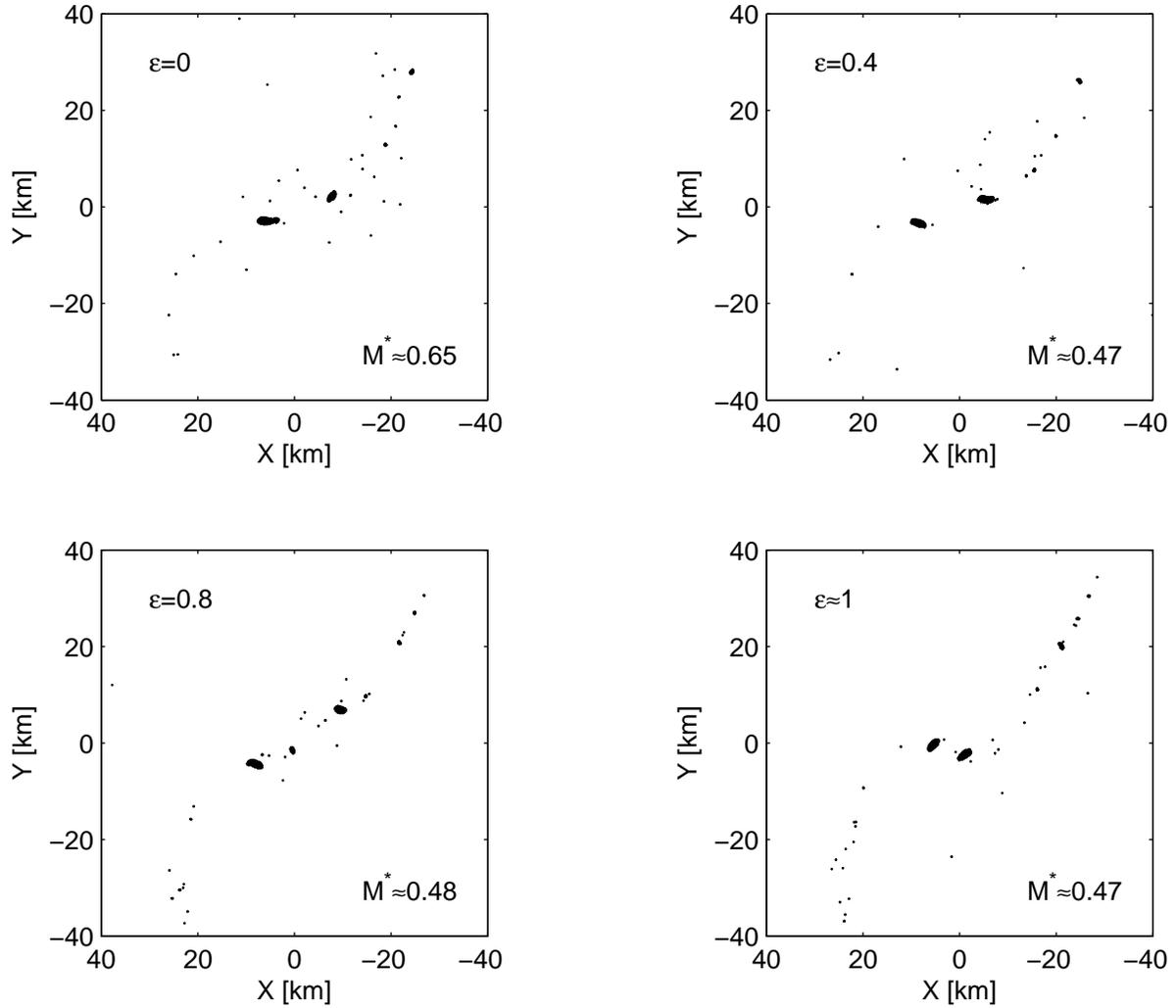}
\caption{Same as Fig.~\ref{fig:fric2} but with varying values of coefficient of restitution. Progenitor rubble piles had a bulk density of $\rho=600\unit{kg/m^3}$. A larger coefficient of restitution speeds up the re-aggregation of clumps after tidal breakup, but by the time clumping is complete the final configuration is not sensitive to the exact value of $\eps$.}
\label{fig:rest}
\end{figure}

The restitution coefficient between cometary material grains in vacuum -- that is, the ratio of the outgoing normal component of velocity, to the incoming normal component of velocity, corresponding to the height that a ball will bounce -- is not known any better than the friction coefficient \citep[see however][]{Hartman1978,Durda2011}. Fortunately our numerical experiments indicate that the exact value of the restitution coefficient is likewise not a main concern (Fig.\ref{fig:rest}), unless the value is extraordinarily high (perfectly elastic collisions) or extraordinarily low (perfectly inelastic, such as might be the case of extremely porous materials). The value of the restitution coefficient does play a role in the timing of the reaccumulaton, but does not affect the timing or extent of breakup, or the number of stable clumps after the encounter is over. Consistent with a number of experimental reports found in the literature \citep{Imre2008,Durda2011} we use a value of $\eps=0.8$ for most of our runs.  But as seen in the figure, even perfectly inelastic collisions are morphologically similar to the $\eps=0.8$ collisions.  Thus, in summary, we find that using nominal values for $\mu$ and $\eps$ to be a justified approach in greatly reducing the parameter space.

\subsubsection{Spin rate} There is no doubt that a rotating progenitor would exhibit a very different behavior, for a given orbit and density, than a non-rotating one. In fact, since the centrifugal acceleration would be added on top of all other forces, it is logical to expect that density and spin rate would be almost complementary. In other words, a range of disruption levels can be achieved either by adjusting density for a given spin rate, or by adjusting spin rate for a given density \citep{Asphaug1996}. Unfortunately, there is no information on SL9's rotation state prior to its breakup, other than statistical knowledge that Jupiter family comets have rotation periods ranging from $\about{6}$ hours to $\about{2}$ days, mostly in the slower range, regulated by their levels of activity. Furthermore there is no knowledge whether this is prograde rotation or retrograde rotation, or out of the plane.  Thus, our results below apply only to a non-rotating comet.  It is important to recognize that retrograde rotation can prevent the nucleus from disrupting tidally, forming a massive central clump that was not observed for SL9.  Thus \cite{Asphaug1996} considered only zero rotation or prograde rotation, finding that a fast-spinning prograde nucleus (period 9~h) would come apart similarly, even if it were $\about{2/3}$ the predicted diameter as a non-rotating nucleus, and $\about{3/2}$ the bulk density.  We do not consider rotation in our present analysis, but anticipate similarly, that what we obtain below are lower limits to the bulk density, given the possibility (relatively minor) that SL9 could have been a prograde rapid rotator.  

\subsubsection{The effect of `resolution'}
In DEM models, the number of discrete, elementary particles is often considered the resolution of the model. This is appropriate in the sense that increasing this number as much as computation resources allow is desirable, and probably leads to more physical results, in the sense that asteroids and comets could be made of relatively fine materials. It is important, however, to realize that this `resolution' has a different meaning than what is usually ascribed to this parameter in numerical continuum models (such as hydrocodes). In a grid-based finite difference scheme, say, the goal is to approximate a continuous field by its value at a finite number of grid points, and the quality of the approximation can be rigorously related to the grid spacing -- the resolution. A natural way to decide what is an appropriate resolution is to look for convergence of the predicted field value with successive halving of the grid spacing.

A DEM is not usually amenable to a rigorous error analysis. We do not have an upper bound to the expected error and no theoretical basis for selecting an appropriate resolution, unless -- as in our laboratory experiment of 1400 12-sided dice falling to their angle of repose described above -- we know it apriori. Unless we are modeling a very small system (or a small volume in a periodic system) we often must use a much smaller number of particles than exists in the physical system. In this work, for example, we use  thousands of grains to model a rubble pile who must in reality contain billions of grains. And we have no theoretical estimate to guide us when trying to understand how good of an approximation we are making.

\begin{figure}
\centering
\includegraphics[width=\textwidth]{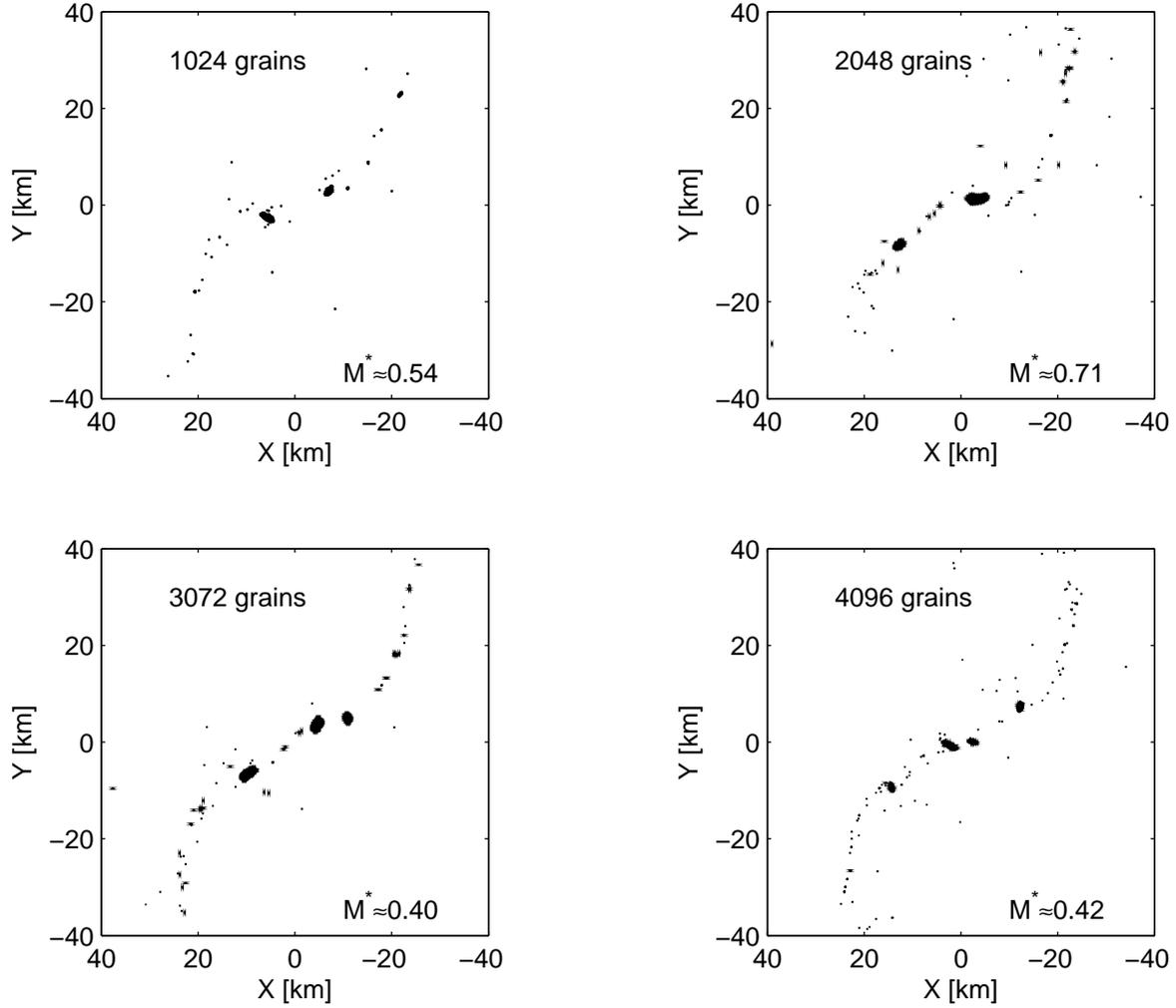}
\caption{Same as Fig.~\ref{fig:rest} but with different number of grains in the progenitor rubble pile. The rubble piles had a bulk density of $\rho=600\unit{kg/m^3}$. The general ensemble behavior is captured almost as well with 1024 grains as with 4096 grains. Whether or not either is a good approximation to an ensemble of billions of grains remains open.}
\label{fig:reso}
\end{figure}

Worse, we cannot even expect that using more and more grains will make our simulations converge, in the usual sense. Figure~(\ref{fig:reso}) shows the result of a tidal encounter modeled at different resolutions. We do not see, and should not expect to see, a convergence in the positions of individual grains, or even of the largest clumps. Although we wish to model the same physical system (the comet) with increasingly better approximations, we are in fact modeling four quite different bodies. Indeed, to keep the mean density constant we must assign slightly different material densities to each of the rubble piles, as their bulk volume cannot be made exactly identical.

We can, however, expect a convergence of sorts with successive increase in resolution. We can expect that the general behavior of the ensemble of grains will not depend on the number of grains, in the sense that a pile of gravel pouring out of a dump truck, will behave similarly to a dump truck load full of sand, with similar total mass, grain size and shape and porosity, but with millions of times more grains.  But a dump truck load of microfine powder will begin to behave differently, as will a dump truck containing four or five huge rocks.  All we can  expect, and require, is that any conclusions we draw from DEM simulations not depend too sensitively on the sizes of the elements. This is indeed the case with our SL9 tidal encounter simulations. The range of bulk densities we find necessary to match the appearance of post-periapse SL9 remains the same when we run our simulations with 1024 grains as when we ran it with 4096 grains. In this sense we believe we have used adequate resolution, a belief that must be retested when computation resources allow.

\subsubsection{Grain geometry} As mentioned above, all previous studies of SL9 using DEM techniques has been done with spherical elements. Our DEM is capable of implementing arbitrary shapes, so another dimension of the parameter space opens up -- what shapes to use? It seems the logical answer is to use random, angular shapes, since there is no reason to think planetary rubble pile elements would have a special shape.  However there is some expectation that fragments generated by past cratering and disruption events will have an average aspect ratio $2:\sqrt{2}:1$ \citep{Fujiwara1989}. All our runs so far were made with 20 sided polyhedra, generated by a prescription that avoids sharp ``knife-edges'' and extreme aspect ratios but is otherwise random.

\begin{figure}
\centering
\includegraphics[width=\textwidth]{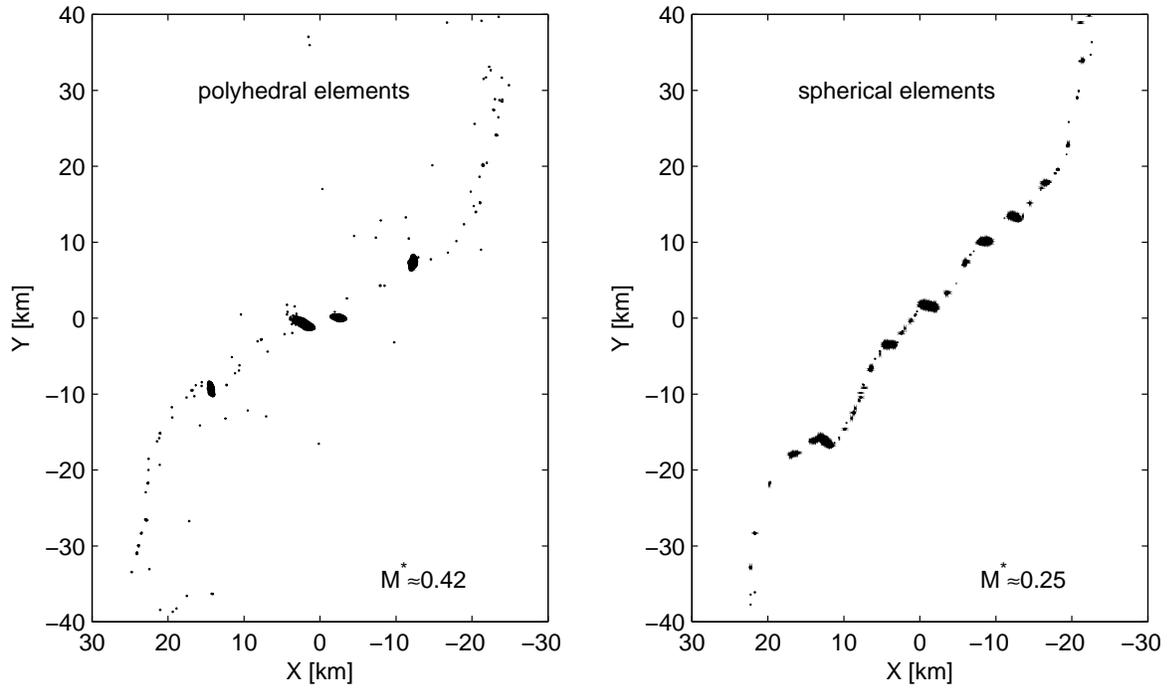}
\caption{Following the same orbit, a rubble pile made of spherical grains behaves differently than a pile made of polyhedral grains. The onset of deformation occurs sooner with spherical grains, and the resulting fragment train contains a larger number of smaller mass fragments. This is probably due to the more pronounced friction and dilatation experienced by polyhedral grains.}
\label{fig:sphr}
\end{figure}

A related question is just how important is the use of non-spherical grains. This is an important question since computer codes that employ only spherical grains can typically run at much higher resolutions, making them perhaps the better choice. We find this choice to be quite important.  We have compared runs with rubble piles composed of spherical grains (using the same \code{PhysX} engine) to runs with similar rubble piles (same shape, same bulk density) made out of random polyhedra. We find that the outcome of a simulated tidal disruption can be significantly different in the two cases (Fig.~\ref{fig:sphr}). Spherical grains appear to be adequate for very ``dynamic'' encounters, where the progenitor is quickly disrupted and the outcome is determined by the play of long range gravity forces between grains. But for gentler encounters, where friction and dilatation forces nearly balance the tidal stress (that is, the kinds of threshold breakups that probably dominate such events), the use of polyhedral grains can make enough of a difference to keep a rubble pile from disrupting at all.  

In future work studying the more general aspects of rubble-pile disruption, we shall focus on the specifics of the aspect ratios of the polyhedra, as these are closely related to what we perceive to be the dominant factors that distinguish the present model from the previous sphere-based models of SL9, namely the physics of grain locking and shear dilatation.

\subsection{\label{sec:mainresult}Progenitor's bulk density}
A tidal encounter between a cohesionless rubble pile and a planet is a tug-of-war game between the gravitational pull of the primary planet, and the self-gravity of the interloping body. The self gravity has three expressions, the most obvious being the long range force that holds particles together. But intergranular friction also scales with the local normal stress, and thus can be considered a result of self-gravity. The third expression is dilatation, the fact that irregular grains, while rotating locally in response to shear, most move other bodies aside, posisbly against the local normal stress. So again, dilatation can be considered an expression of self-gravity.

It is therefore ultimately the rubble pile's bulk density that will  determine the outcome of the encounter (or more precisely, the ratio of the rubble pile's density to the density of the primary, to the $1/3$ power), for a given structural configuration.  Conversely, knowing that the outcome of SL9's encounter with Jupiter was a train of sizable fragments, we may use this to bracket the possible density values of the progenitor rubble pile.

\begin{figure}
\centering
\includegraphics[width=\textwidth]{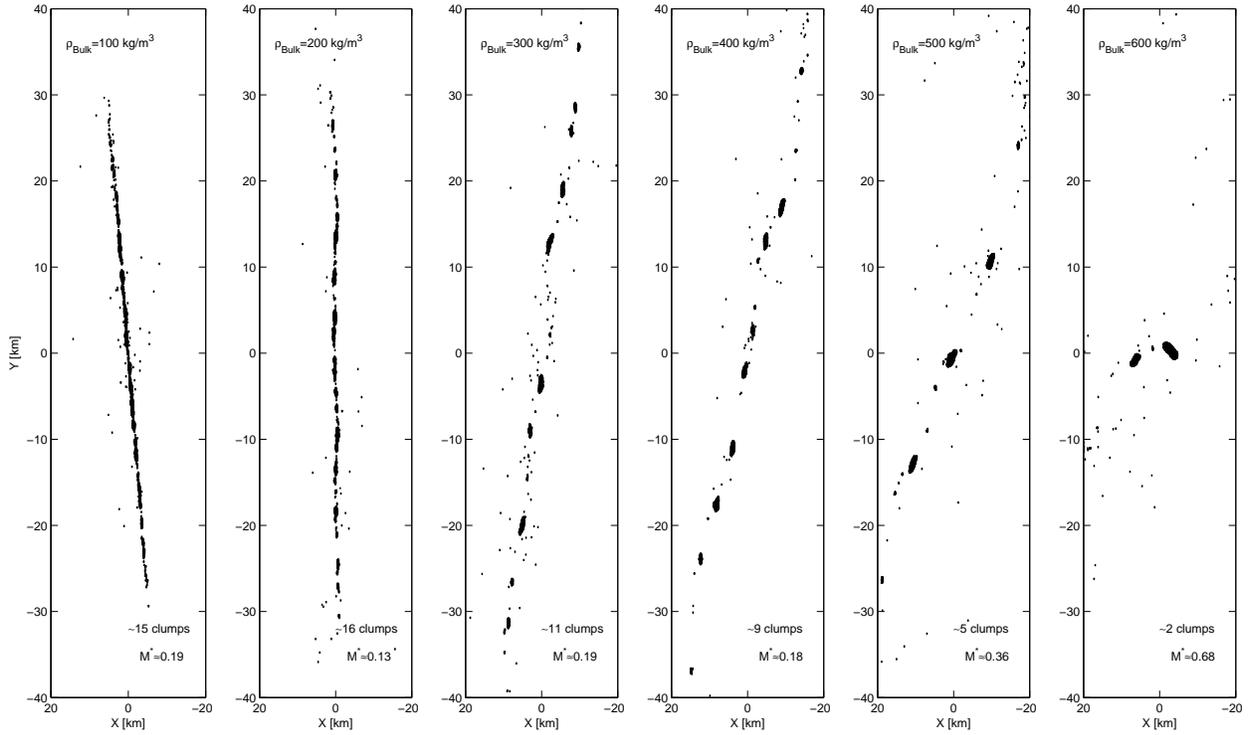}
\caption{Positions of mass fragments, in the orbit plane, of a rubble-pile body following SL9's 1992 orbit, shown $\about{5}\unit{hours}$ after perijove. For these runs a friction coefficient $\mu=0.5$ and restitution coefficient $\eps=0.8$ were used. The lowest and highest bulk density values are wholly inconsistent with the appearance of the SL9 chain, thus providing a revised constraint on the density of this comet.}
\label{fig:sl9dens}
\end{figure}

Using nominal values for the coefficients of friction and restitution, we ran simulated tidal encounters of a non-rotating, $1\unit{km}$ rubble pile of 4096 randomly shaped polyhedra, along SL9's 1992 nearly-parabolic orbit, varying the comet's bulk density from $100$ to $600\unit{kg\;m^{-3}}$. Figure~\ref{fig:sl9dens} shows how varying the bulk density  affects the outcome of the encounter. We can see that a density smaller than $\about{300}\unit{kg\;m^{-3}}$, or greater than $\about{400}\unit{kg\;m^{-3}}$ would be inconsistent with the fragment train observed in the case of SL9, or expected from the appearance of linear crater chains on Ganymede.

\section{Conclusions}

The tidal disruption of comet Shoemaker-Levy~9 has been revisited using state of the art computational techniques for the behavior of  rubble piles.  The primary difference between these new models and what has been done in previous efforts \citep{Asphaug1994,Asphaug1996,Richardson1998} is the use of non-spherical, and thus presumably more realistic grain shapes, so that instead of `marble piles' we have true rubble piles with grains that can lock and jam and pack differently than spheres, and which must do work in order to achieve any amount of shear strain or even rotate in place. In this sense, dilatancy can be considered a force resisting tidal sheer, and this force may well have been underestimated when spherical grains were used to model a rubble pile, for in that case individual grains can respond to sheer by freely rotating in-place, unless measures are taken to explicitly constrain them.

While we include coefficients of friction and coefficients of restitution in our simulations, we find that varying these widely, but within reasonable values, does little to affect the final outcome of tidal disruption. These parameters do affect the timing of disruption and re-aggregation however. Increasing the coefficient of friction allows the progenitor body to better resist tidal sheer, and so the point of runaway deformation and disruption is postponed. Increasing the coefficient of restitution, the elasticity of individual grains, causes the re-aggregation phase to continue for a longer time, as grains bounce off of each other and may take hours to settle in stable clumps.

These observations, however, are unlikely to be the last word on the topic, because friction as implemented is a very simple force, acting when contact surfaces shear past one another at the resolution scale of tens of meters.  For much smaller-scale assemblages (piles of dust, perhaps) internal friction can be much greater.  That said, the overall morphology of the clumps from Shoemaker-Levy 9 are self-gravitational in signature.  This being one example, \cite{Schenk1996} looked to the much larger record of a dozen or more tidal disruption catenae imprinted on the satellites of Jupiter.  Specifically they found no correlation between the number of clumps in a chain, versus the diameter of the progenitor forming the chain.  In other words, the clumping process is scale invariant, suggesting self-gravitation rather than fragmentation, which does depend on scale, is the dominant process.  As for coefficient of restitution, we find even less sensitivity to this parameter, with the range of outcomes being similar  within the range of values that might be applicable.

A bulk density $\about{300}$--${400}\unit{kg\;m^{-3}}$ is obtained for Shoemaker-Levy~9.  This is in good agreement with the recent estimates of cometary bulk density from the Deep Impact mission (highly uncertain), estimated from the rate of fallback of crater ejecta \citep{Richardson2007}, and from the dynamics of orbiting blocks that were imaged by the same spacecraft when it flew by Comet Hartley 2 \citep{AHearn2011}; further constrained by the comet's dog-bone shape. (\citeauthor{AHearn2011} found that $\rho=220\unit{kg\;m^{-3}}$ works best when fitting potential contours to the geometry of the waist.)  This latter result advises caution in blindly adopting the physical realism of ours or any other forward model, because it shows that even our present treatment, however technically advanced, does not yet address such factors as shape and initial rotation.

That said, it is important to study these rare events because they are the only tests we shall have, at a cosmic-geological scale, of global scale catastrophic disruption of small bodies.  It is the only way, pending spacecraft investigations that activate a comet nucleus globally (e.g.~seismology) or that form mega-craters on small asteroids (precursors to hazard mitigation, or for science), that we are going to be able to learn how rubble piles behave, and evolve, and respond to energetic impulses and planetary encounters.   By advancing the predictive capability of rubble-pile models, and  benchmarking them against  known small bodies data, we are making forward predictions more reliable.

\acknowledgments
The authors wish to thank Derek Richardson for helpful suggestions during the review process. This effort was sponsored by NASA's Near Earth Object Observations Program, award number NNX10AG52G, and by NASA's Planetary Geology and Geophysics Program, award number NNX07AQ04G.

\bibliographystyle{apj}

\begin{thebibliography}{46}
\expandafter\ifx\csname natexlab\endcsname\relax\def\natexlab#1{#1}\fi

\bibitem[{A'Hearn {et~al.}(2011)A'Hearn, Belton, Delamere, Feaga, Hampton,
  Kissel, Klaasen, McFadden, Meech, Melosh, Schultz, Sunshine, Thomas, Veverka,
  Wellnitz, Yeomans, Besse, Bodewits, Bowling, Carcich, Collins, Farnham,
  Groussin, Hermalyn, Kelley, Li, Lindler, Lisse, McLaughlin, Merlin,
  Protopapa, Richardson, \& Williams}]{AHearn2011}
A'Hearn, M., Belton, M. J.~S., Delamere, W.~A., {et~al.} 2011, Science (New
  York, N.Y.), 332, 1396

\bibitem[{Albert {et~al.}(1997)Albert, Albert, Hornbaker, Schiffer, \&
  Barab\'{a}si}]{Albert1997}
Albert, R., Albert, I., Hornbaker, D., Schiffer, P., \& Barab\'{a}si, A.-L.
  1997, Physical Review E, 56, R6271

\bibitem[{Asphaug(2009)}]{Asphaug2009a}
Asphaug, E. 2009, Annual Review of Earth and Planetary Sciences, 37, 413

\bibitem[{Asphaug \& Benz(1994)}]{Asphaug1994}
Asphaug, E., \& Benz, W. 1994, Nature, 370, 120

\bibitem[{Asphaug \& Benz(1996)}]{Asphaug1996}
Asphaug, E., \& Benz, W. 1996, Icarus, 121, 225

\bibitem[{Belton {et~al.}(2004)Belton, Morgan, Samarasinha, \&
  Yeomans}]{Belton2004}
Belton, M. J.~S., Morgan, T.~H., Samarasinha, N.~H., \& Yeomans, D. 2004, in
  Mitigation of Hazardous Comets and Asteroids

\bibitem[{Benz \& Asphaug(1994)}]{Benz1994}
Benz, W., \& Asphaug, E. 1994, Icarus, 107, 98

\bibitem[{Benz \& Asphaug(1995)}]{Benz1995}
Benz, W., \& Asphaug, E. 1995, Computer physics communications, 87, 253

\bibitem[{Boss(1994)}]{Boss1994}
Boss, A. 1994, Icarus, 107, 422

\bibitem[{Bottke \& Melosh(1996)}]{Bottke1996a}
Bottke, W.~F., \& Melosh, H. 1996, Icarus, 124, 372

\bibitem[{Bottke {et~al.}(1997)Bottke, Richardson, Love, \&
  Others}]{Bottke1997}
Bottke, W.~F., Richardson, D.~C., Love, S.~G., \& Others. 1997, Icarus, 126,
  470

\bibitem[{Chandrasekhar(1961)}]{Chandrasekhar1961}
Chandrasekhar, S. 1961, {Hydrodynamic and hydromagnetic stability}, ed.
  S.~Chandrasekhar (Oxford University Press)

\bibitem[{Chodas \& Yeomans(1996)}]{Chodas1996}
Chodas, P., \& Yeomans, D. 1996, in The Collision of Comet Shoemaker-Levy 9 and
  Jupiter, ed. K.~Noll, H.~Weaver, \& P.~Feldman (Cambridge: Cambridge Univ.
  Press), 1--30

\bibitem[{{Committee to Review Near-Earth Object Surveys and Hazard Mitigation
  Strategies Space Studies Board}(2010)}]{NRC2010}
{Committee to Review Near-Earth Object Surveys and Hazard Mitigation Strategies
  Space Studies Board}. 2010, {Defending Planet Earth:Near-Earth Object Surveys
  and Hazard Mitigation Strategies} (The National Academies Press)

\bibitem[{Crawford {et~al.}(1995)Crawford, Boslough, Trucano, \&
  Robinson}]{Crawford1995}
Crawford, D., Boslough, M., Trucano, T., \& Robinson, A. 1995, International
  journal of impact engineering, 17, 253

\bibitem[{Durda {et~al.}(2011)Durda, Movshovitz, Richardson, Asphaug, Morgan,
  Rawlings, \& Vest}]{Durda2011}
Durda, D.~D., Movshovitz, N., Richardson, D.~C., {et~al.} 2011, Icarus, 211,
  849

\bibitem[{Fujiwara {et~al.}(1989)Fujiwara, Cerroni, Davis, Ryan, \&
  di~Martino}]{Fujiwara1989}
Fujiwara, A., Cerroni, P., Davis, D., Ryan, E., \& di~Martino, M. 1989, in
  Asteroids II, 240--265

\bibitem[{Fujiwara {et~al.}(2006)Fujiwara, Kawaguchi, Yeomans, Abe, Mukai,
  Okada, Saito, Yano, Yoshikawa, Scheeres, Barnouin-Jha, Cheng, Demura,
  Gaskell, Hirata, Ikeda, Kominato, Miyamoto, Nakamura, Nakamura, Sasaki, \&
  Uesugi}]{Fujiwara2006}
Fujiwara, A., Kawaguchi, J., Yeomans, D., {et~al.} 2006, Science (New York,
  N.Y.), 312, 1330

\bibitem[{Gilbert {et~al.}(1988)Gilbert, Johnson, \& Keerthi}]{Gilbert1988}
Gilbert, E., Johnson, D., \& Keerthi, S. 1988, IEEE Journal on Robotics and
  Automation, 4, 193

\bibitem[{Hahn \& Rettig(1998)}]{Hahn1998a}
Hahn, J., \& Rettig, T. 1998, Planetary and Space Science, 46, 1677

\bibitem[{Hartman(1978)}]{Hartman1978}
Hartman, W. 1978, Icarus, 33, 50

\bibitem[{Holsapple \& Michel(2006)}]{Holsapple2006}
Holsapple, K.~A., \& Michel, P. 2006, Icarus, 183, 331

\bibitem[{Holsapple \& Michel(2008)}]{Holsapple2008}
Holsapple, K.~A., \& Michel, P. 2008, Icarus, 193, 283

\bibitem[{Imre {et~al.}(2008)Imre, R\"{a}bsamen, \& Springman}]{Imre2008}
Imre, B., R\"{a}bsamen, S., \& Springman, S. 2008, Computers \& Geosciences,
  34, 339

\bibitem[{Jeans(1917)}]{Jeans1917}
Jeans, J. 1917, in Mem. R. Astron. Soc. London 62, 1--48

\bibitem[{Knight {et~al.}(2010)Knight, A'Hearn, Biesecker, Faury, Hamilton,
  Lamy, \& Llebaria}]{Knight2010}
Knight, M., A'Hearn, M., Biesecker, D., {et~al.} 2010, The Astrophysical
  Journal, 139, 926

\bibitem[{Korycansky \& Asphaug(2009)}]{Korycansky2009}
Korycansky, D., \& Asphaug, E. 2009, Icarus, 204, 316

\bibitem[{Longshaw {et~al.}(2009)Longshaw, Turner, Finch, \&
  Gawthorpe}]{Longshaw2009}
Longshaw, S.~M., Turner, M.~J., Finch, E., \& Gawthorpe, R. 2009, EG UK Theory
  and Practice of Computer Graphics

\bibitem[{MacMillan(1936)}]{MacMillan36}
MacMillan, W.~D. 1936, {Dynamics of Rigid Bodies} (New York, New York, USA:
  Dover Publications), 478

\bibitem[{Marsden(1967)}]{Marsden1967}
Marsden, B.~G. 1967, The Astronomical Journal, 72, 1170

\bibitem[{Melosh \& Whitaker(1994)}]{Melosh1994}
Melosh, H., \& Whitaker, E. 1994, Nature, 369, 713

\bibitem[{Merline {et~al.}(2002)Merline, Weidenschilling, Durda, Margot,
  Pravec, \& Storrs}]{Merline2002}
Merline, W.~J., Weidenschilling, S., Durda, D.~D., {et~al.} 2002, in Asteroids
  III, 289--312

\bibitem[{Noll {et~al.}(1996)Noll, Weaver, \& Feldman}]{Noll1996}
Noll, K., Weaver, H., \& Feldman, P. 1996, in IAU Colloq. 156: The Collision of
  Comet Shoemaker-Levy 9 and Jupiter, ed. K.~Noll, H.~Weaver, \& P.~Feldman

\bibitem[{Richardson {et~al.}(1998)Richardson, Bottke, \&
  Love}]{Richardson1998}
Richardson, D.~C., Bottke, W.~F., \& Love, S.~G. 1998, Icarus, 134, 47

\bibitem[{Richardson {et~al.}(2002)Richardson, Leinhardt, Melosh, Bottke, \&
  Asphaug}]{Richardson2002}
Richardson, D.~C., Leinhardt, Z.~M., Melosh, H., Bottke, W.~F., \& Asphaug, E.
  2002, in Asteroids III, ed. W.~F. Bottke, A.~Cellino, P.~Paolocchi, \&
  R.~Binzel (Tucson: University of Arizona Press), 501--515

\bibitem[{Richardson {et~al.}(2007)Richardson, Melosh, Lisse, \&
  Carcich}]{Richardson2007}
Richardson, J., Melosh, H., Lisse, C., \& Carcich, B. 2007, Icarus, 190, 357

\bibitem[{Roche(1847)}]{Roche1847}
Roche, E. 1847, in Acad. Sci. Lett. Montpelier. Mem. Section Sci. 1, 243--262

\bibitem[{Schenk {et~al.}(1996)Schenk, Asphaug, McKinnon, Melosh, \&
  Weissman}]{Schenk1996}
Schenk, P., Asphaug, E., McKinnon, W., Melosh, H., \& Weissman, P. 1996,
  Icarus, 121, 249

\bibitem[{Schwartz {et~al.}(2012)Schwartz, Richardson, \&
  Michel}]{Schwartz2012}
Schwartz, S.~R., Richardson, D.~C., \& Michel, P. 2012, Granular Matter, 14,
  363

\bibitem[{Sekanina {et~al.}(1994)Sekanina, Chodas, \& Yeomans}]{Sekanina1994}
Sekanina, Z., Chodas, P., \& Yeomans, D. 1994, Astronomy and Astrophysics, 289,
  607

\bibitem[{Solem(1994)}]{Solem1994}
Solem, J. 1994, Nature, 370, 349

\bibitem[{Solem \& Hills(1996)}]{Solem1996}
Solem, J., \& Hills, J.~G. 1996, The Astronomical Journal, 111, 1382

\bibitem[{Sridhar \& Tremaine(1992)}]{Sridhar1992}
Sridhar, S., \& Tremaine, S. 1992, Icarus, 95, 86

\bibitem[{Weaver {et~al.}(1995)Weaver, A'Hearn, Arpigny, Boice, Feldman,
  Larson, Lamy, Levy, Marsden, \& Meech}]{Weaver1995}
Weaver, H., A'Hearn, M., Arpigny, C., {et~al.} 1995, Science, 267, 1282

\bibitem[{Weidenschilling {et~al.}(1989)Weidenschilling, Paolicchi, \&
  Zappala}]{Weidenschilling1989}
Weidenschilling, S., Paolicchi, P., \& Zappala, V. 1989, in Asteroids II,
  643--658

\bibitem[{Weissman {et~al.}(2004)Weissman, Asphaug, \& Lowry}]{Weissman2004}
Weissman, P., Asphaug, E., \& Lowry, S. 2004, in Comets II, ed. M.~Festou,
  H.~Keller, \& H.~Weaver (University of Arizona Press), 337--357

\end{thebibliography}

\end{document}